\pdfoutput=1

\documentclass[final,3p,times,twocolumn]{elsarticle}

\usepackage{graphicx}

\usepackage{amssymb}

\usepackage{lineno}

\biboptions{sort&compress}

\journal{Astroparticle Physics}

\begin{document}

\begin{frontmatter}


\title{A Search for the Dark Matter Annual Modulation in South Pole Ice}


\author[16]{J.~Cherwinka}

\author[3]{R.~Co\fnref{fn1}}

\author[9,8]{D.~F.~Cowen}

\author[1]{D.~Grant}

\author[14]{F.~Halzen}

\author[14]{K.~M.~Heeger}

\author[3]{L.~Hsu\corref{cor1}}
\ead{llhsu@fnal.gov}

\author[14]{A.~Karle}

\author[11]{V.~A.~Kudryavtsev}

\author[14]{R.~Maruyama}

\author[14]{W.~Pettus}

\author[11]{M.~Robinson}

\author[11]{N.~J.~C.~Spooner}

\cortext[cor1]{Corresponding author}
\fntext[fn1]{Present Address: Physics Department, University of California, Berkeley, CA 94720, USA}

\address[16]{Physical Sciences Laboratory, University of Wisconsin, Stoughton WI 53589, USA}
\address[3]{Fermi National Accelerator Laboratory, Batavia, IL 60510, USA}
\address[9]{Department of Astronomy and Astrophysics, Pennsylvania State University, University Park, PA 16802, USA}
\address[8]{Department of Physics, Pennsylvania State University, University Park, PA 16802, USA}
\address[1]{Department of Physics, University of Alberta, Edmonton, Alberta, Canada}
\address[14]{Department of Physics, University of Wisconsin, Madison, WI 53706, USA}
\address[11]{Department of Physics and Astronomy, University of Sheffield, Sheffield S3 7RH, UK}



\begin{abstract}
Astrophysical observations and cosmological data have led to the conclusion that nearly one quarter of the Universe consists of dark matter. Under certain assumptions, an observable signature of dark matter is the annual modulation of the rate of dark matter-nucleon interactions taking place in an Earth-bound experiment. To search for this effect, we introduce the concept for a new dark matter experiment using NaI scintillation detectors deployed deep in the South Pole ice. This experiment complements dark matter search efforts in the Northern Hemisphere and will investigate the observed annual modulation in the DAMA/LIBRA and DAMA/NaI experiments.  The unique location will permit the study of background effects correlated with seasonal variations and the surrounding environment.  This paper describes the experimental concept and explores the sensitivity of a 250\,kg NaI experiment at the South Pole.
\end{abstract}


\end{frontmatter}


\newcommand{\gev}{\ensuremath{{\rm GeV}/{\rm c}^2}}
\newcommand{\smod}{\ensuremath{{S_m}}}
\newcommand{\snaught}{\ensuremath{{S_0}}}
\newcommand{\rzero}{\ensuremath{{R_0}}}
\newcommand{\nzero}{\ensuremath{{N_0}}}
\newcommand{\keVee}{\ensuremath{{\rm keV_{ee}}}}
\newcommand{\degree}{\ensuremath{{^{\circ}}}}



\section{Introduction}
\label{sec:intro}

An abundance of astrophysical and cosmological observations have established a modern concordance model of cosmology, known as $\Lambda$CDM~\cite{lamdacdm}.  This model prescribes that 23\% of the Universe can be attributed to cold dark matter, 4\% to baryonic matter and the remaining 73\% to dark energy. Much of what we know about dark matter is inferred solely from its gravitational interactions.  While evidence for dark matter has been firmly established~\cite{DMreview}, dark matter has yet evaded direct observation and thus many of its aspects remain speculative. 

Weakly interacting massive particles (WIMPs) are a theoretically favored candidate for dark matter~\cite{WIMPs}.  A suite of direct detection experiments is now underway worldwide to search for WIMPs through observation of WIMP-nucleon elastic scattering.   The sensitivity of these experiments is now sufficient to allow interesting theoretical regimes to be probed, such as those predicted by the Minimal Supersymmetric Model~\cite{Jungman:1995df}.

A WIMP moving at typical galactic rotation speeds of $\sim$220\,km/s could in principle interact coherently with a nucleus, inducing a nuclear recoil with energy of a few tens of keV~\cite{GoodmanWitten}.  Detecting such interactions requires stringent control of background and long, stable exposure times in low-noise environments.  This can be achieved by operating detectors in ultra-low-radioactivity environments and at underground locations to minimize activity induced by cosmic rays.  Many such detectors make use of phenomena by which the particle interaction energy is observed through more than one channel (e.g. phonon and ionization), the ratio of which allows nuclear recoil events to be distinguished from an electron recoil background~\cite{directDMReview}.

An additional strategy is to search for a modulation in the detected event rate.   Under commonly accepted models, galactic dark matter is distributed in a diffuse, roughly spherical halo through which the solar system travels as the Sun moves within the disk of the Milky Way\cite{DMhalos}.  The rate of events detected by an Earth-bound experiment depends on the relative velocity of the dark matter to the detecting medium.  The rotation of the Earth about the Sun produces a small but periodic variation in this relative velocity.   This affects the rate of dark matter interactions in the earth-bound detector and the energy spectrum of the induced recoils.  If the coupling to baryonic matter is large enough, a periodic change in the rate of particle interactions may be observable~\cite{PhysRevD.33.3495}.  Such a modulation would exhibit a period of one Earth-year and peak at the time of year when the Earth-dark matter relative velocity is at an extremum.  

The DAMA/NaI experiment and its successor, DAMA/LIBRA, have been operating an array of sodium iodide (NaI) crystals in the Laboratori Nazionali del Gran Sasso (LNGS).  They have reported a statistically significant observation of an annual modulation, which has been attributed to dark matter~\cite{Bernabei:2010mq,damafullpaper}.  However, the interpretation of these data as evidence for elastic scattering of WIMPs off sodium (Na) or iodine (I) nuclei remains inconsistent with the observations from a number of direct detection experiments~\cite{xenon2011, cdms2010, edelweissii, coupp2011, picasso, zepliniii, NAIAD}.  Moreover, the presence of a signal below 6 keV, as claimed by DAMA, would require a significant suppression of background continuum compared to energies above 6 keV, in contradiction with Monte Carlo simulations of radioactive background~\cite{vak2010}.  


The challenge of reconciling the DAMA observations with those of other experiments has generated a great deal of interest in alternative models of dark matter (e.g.~\cite{inelastic,ivdm,momdepdm}).  To date, none of the proposed scenarios has satisfactorily accounted for the DAMA modulation.  It remains that the DAMA claim has never been tested at an independent location with the same detector target material (NaI) and an experimental setup with equivalent sensitivity.  Given these motivations, we describe an annual modulation search using NaI crystals that can be deployed in the South Pole ice.  


%


\section{Annual modulation search at the South Pole}
\label{sec:methods}

\subsection{Conceptual Description}
\label{sec:concept}


This paper describes a NaI dark matter experiment that may be deployed to a depth of 2450\,m in the Antarctic icecap.  This Southern Hemisphere location provides a unique opportunity to disentangle seasonal effects from an annual modulation caused by dark matter.  The ice offers a quiet and stable operating environment, a necessary requirement for an annual modulation search.  The proposed location is at the bottom of the IceCube neutrino detector.   Section~\ref{sec:concept} provides an overview of the basic design and requirements for such an experiment.  It also describes the way in which the suggested location meets these goals.   Details of background estimates and sensitivity are provided in subsequent sections.

An annual modulation arising from dark matter will have the same phase regardless of the location on Earth. Seasonal effects, however, are opposite in the Northern and Southern Hemispheres.  For instance, at the South Pole, the peak of a dark matter signal that is consistent with the DAMA result will occur six months out of phase from the peak in cosmogenic muon flux and associated background.  Such a location thus allows one to disentangle seasonal environmental effects from astrophysical effects.   Further, the depth of ice provides a significant overburden that is comparable to underground laboratories, significantly reducing the flux of cosmogenic muons passing through or near the dark matter detectors.  Data from IceCube provide detailed measurements of the muon flux.  Correlations with these data may be exploited to remove any remaining cosmogenic background from a potential dark matter signal (Section~\ref{sec:cosmo_bg}).  

An experiment consisting of 250\,kg of ultra-pure NaI crystal scintillators, which is comparable in size to DAMA/LIBRA, can confirm or exclude the DAMA dark matter claim with a minimum of two years operation time (Section~\ref{sec:sensitivity}).  Key to such an effort is the development of crystals that have a comparable or lower internal radioactive contamination than the DAMA NaI crystals.  The DAMA/LIBRA experiment observed a single-scatter event rate of $\sim$1 count per day (cpd) per kg per \keVee~(electron-recoil equivalent energy) in the 2-10\,\keVee~region of interest~\cite{damafullpaper}.  The observed modulation amounts to a $\sim$2\% change in the rate of events observed between 2-6\,\keVee.  Reproducing such a measurement requires achievement of a very low energy threshold and good control of environmental background.

The Antarctic ice provides a radiopure and stable environment for a dark matter experiment.  Radioactivity levels in Antarctic ice are estimated to be comparable to the cleanest shielding materials used in underground experiments, removing the need for additional shielding from the ice (Section~\ref{sec:radio_bg}).  The level of radon in the ice is expected to contribute a negligible component of the total background.  The temperature of the ice at the South Pole varies in time by less than a degree.  Based on a known profile measured by IceCube, the temperature at the surface is $-50\degree$\,C and increases to $-20\degree$\,C at a depth of 2500\,m.  The measured and calculated profiles of the temperature versus depth at various sites across Antarctica agree to within 0.2\degree~\cite{PriceAge}.  At such temperatures, the light output of a NaI crystal is several percent higher compared to room temperature; however, scintillation will occur over a longer time frame~\cite{stgobain,Ianakiev2006Temperature}.  

With the successful installation and operation of nearly 5200 digital optical modules in $\sim$1\,km$^3$ of ice, IceCube has demonstrated the technical feat of safely deploying delicate instrumentation in Antarctic ice and has achieved reliable, remote operation of these devices.  The scientific infrastructure at the Amundsen-Scott South Pole Station, operated by the National Science Foundation, provides excellent technical and scientific support.  Utilizing much of the technology and expertise developed by IceCube, it is possible to attach encapsulated NaI crystals and their PMT systems to deployment-readout cables similar to those used in the IceCube array.   A variety of deployment depths, up to a maximum of 2500m, are under consideration.

Challenges associated with a deep-ice deployment, although not insurmountable, must be adequately addressed in the experimental design.  Dark matter detectors require a method of robust in-situ calibration.  Possibilities within the ice include LED's, and natural and cosmogenic radioactivity in NaI.  Transport to and storage at the South Pole Station, at an altitude of 2850\,m, will result in cosmic activation of the NaI crystals, readout devices and encapsulation assembly.  Depending on the activated isotopes, this activity will decay away on a variety of timescales and may result in a period of higher background at the start of the data collection period.  Once a detector is deployed, it becomes permanently inaccessible and thus reliability of operation will be critical.   NaI scintillators represent an ideal choice for such a requirement because they are relatively simple to operate and have a history of use in field-based research.  These issues and others are currently under study for optimization in a polar-ice deployment.


\subsection{Background from Environmental Radioactivity}
\label{sec:radio_bg}

Gammas, betas, alphas and neutrons will arise from trace radioactive contaminants in the crystals and the surrounding material (PMTs, pressure vessel, shielding, ice, etc.).  Some of these particles will produce signals in the NaI that are indistinguishable from nuclear recoils expected from WIMPs.  The annual modulation search is a compelling strategy for detectors that cannot distinguish between nuclear and electron recoils.  In such cases, one may search for a modulation in event rate.  As discussed in Section~\ref{sec:sensitivity}, the sensitivity of such an experiment is limited by the level of background that is achieved. 

\begin{figure}[!th]
\begin{center}
\includegraphics[angle=90, width=0.45\textwidth]{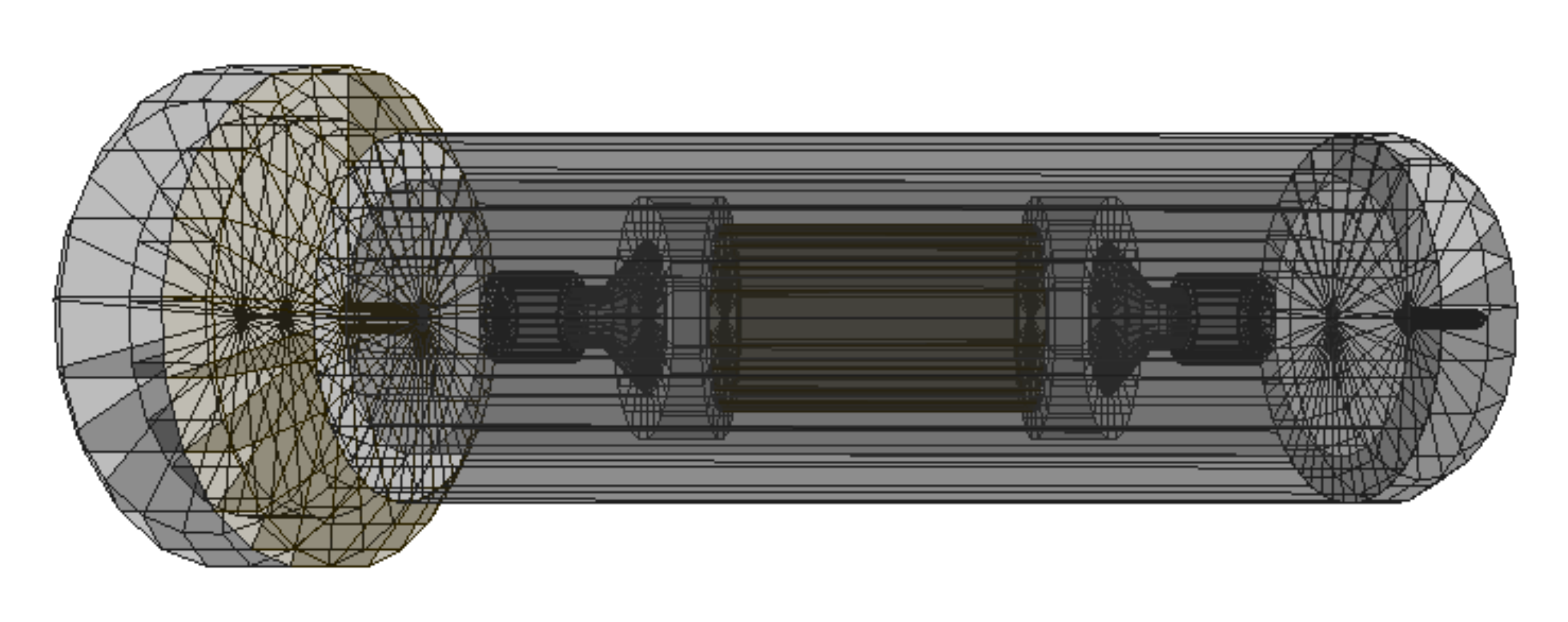} 
\caption{Geant4 visualization of the simple detector assembly used for radioactive background estimates.  Shown is the steel PV (large cylinder with enlarged cap on the left side and rounded bottom on the right side), with an 8\,kg NaI crystal inside (small, interior cylinder).  The crystal is instrumented with a light guide and PMT on each end.}
\label{fig:g4_detector}
\end{center}
\end{figure}

We perform an initial estimate of such background events expected in a 1-10\,\keVee~window.   The estimate consists of a Geant4 simulation~\cite{geant4} of radioactive decay from several primary components of a detector assembly and the surrounding ice.  We consider a unit consisting of a single 8\,kg crystal that is instrumented with two light guides, two PMTs and encased in a stainless steel pressure vessel (PV) as shown in Figure~\ref{fig:g4_detector}.  Typical levels of radioactivity for the assembly materials are input to the simulation and are summarized in Table~\ref{tab:mc_input}.   

\begin{table}[thb]
\caption{Assumed concentrations of $^{238}$U, $^{232}$Th and $^{nat}$K, in ppb for major components of a NaI assembly and the surrounding ice.  Details on the estimate for contamination in Antactic ice are in the text.}
\begin{center}
\begin{tabular}{lccc} 
\hline
Material                      & $^{238}$U & $^{232}$Th & $^{nat}$K \\
\hline
drill ice ~\cite{SNOLAB}      & 0.076$\pm$0.046 & 0.47$\pm$0.14 & $<$262 \\
Antarctic ice                 & $10^{-4}$ & $10^{-4}$  & 0.1 \\
PMT~\cite{ETL}                & 30        & 30         & 60000 \\
steel PV~\cite{SNOLAB}        & 0.2       & 1.6        & 442  \\
NaI                           & 0.005     & 0.005      & 10   \\
\hline
\end{tabular}
\end{center}
\label{tab:mc_input}
\end{table}


In the simulation, the ice surrounding the detector is treated as two distinct volumes, the drill ice and the glacial ice.  In reality, the volume immediately surrounding the pressure vessel is expected to consist of a $\sim$60\,cm diameter column of ``drill'' ice that is formed when hot, drill water mixes with melted glacial ice and refreezes in the drill hole.   Surrounding this hole will be pure, Antarctic glacial ice.  To assess the activity levels expected from the drill ice, samples from several holes were taken during IceCube construction.  Using 2~liter polypropylene bottles, the samples were gathered from the drill return water, which was pumped out from the top of the drill hole.    The samples were counted at SNOLAB~\cite{SNOLAB}.  The results of the count are shown in Table~\ref{tab:mc_input}.  Improvement of the contamination level can be achieved by changing the recirculation of water through the drill.  We simulate the drill ice contamination within a column of ice with a diameter 40\,cm larger than the diameter of the vessel and a length 2\,m longer than the vessel.  The results are shown in Table~\ref{tab:mc_output}.

\begin{figure}[!t]
\begin{center}
\includegraphics[angle=0, width=0.47\textwidth]{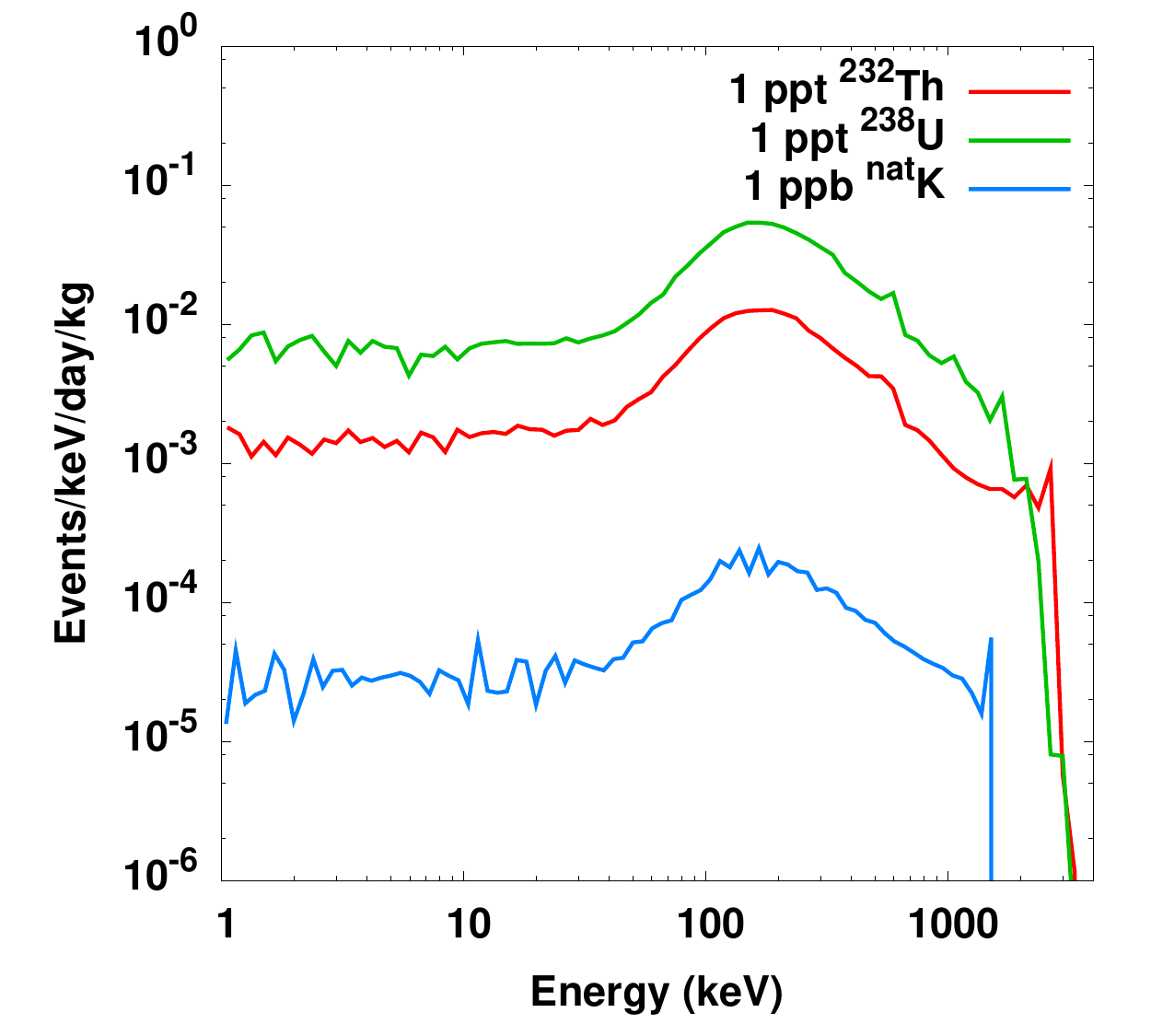} 
\caption{The histograms show the simulated activity in NaI crystals due to uniform radioactivity in the ice surrounding the detector assembly.}
\label{fig:uthk_spectrum}
\end{center}
\end{figure}

The glacial ice background is simulated by assuming the contaminants are uniformly distributed in a spherical volume of ice, 2\,m in radius, surrounding the detector.  Figure~\ref{fig:uthk_spectrum} shows the simulated activity in the NaI crystals arising from contamination in this volume of ice at the level of 1\,ppt $^{238}$U and $^{232}$Th and 1\,ppb of $^{nat}$K.  We scale the activity levels shown in Figure~\ref{fig:uthk_spectrum} to match the contamination levels expected in the Antarctic glacial ice.  Details on the estimated radiopurity of the Antarctic ice are in the next paragraph.   The simulation results are shown in table~\ref{tab:mc_output}.  We note that this calculation overestimates the glacial ice background because we have already accounted for the volume immediately surrounding the pressure vessel with the drill ice simulation; however, the contribution of the glacial ice is nearly negligible so this assumption has little impact on the results.

Although no ice core samples have been extracted at the South Pole, we can estimate the glacial ice contamination levels from the measurements taken at Vostok located 1300\,m away from the South Pole~\cite{vostok}~\cite{Petit:1999fk}.  Most of the contaminants arise from volcanic ash accumulated as dust layers in the ice.  Optical measurements from AMANDA and IceCube allow us to correlate the dust layers present at the South Pole to those at Vostok~\cite{PriceAge}~\cite{Ackermann2006Optical}.  The ice at the South Pole between 2100\,m to 2500\,m deep is roughly 70,000 to 100,000 years old, and contains about 0.1\,ppm of dust.  Volcanic dust typically contains about 1\,ppm of $^{238}$U and $^{232}$Th and 1000\,ppm of $^{nat}$K~\cite{roseBornhorst}, yielding contamination levels of $\sim$10$^{-4}$\,ppb in the ice for $^{238}$U and $^{232}$Th and 0.1\,ppb for $^{nat}$K.    The $^{238}$U concentration was measured independently in the Vostok ice core samples using inductively coupled plasma sector field mass spectrometry (ICP-SFMS), yielding consistent results~\cite{vostok}.

The results of the simulation for all considered materials is summarized in Table~\ref{tab:mc_output}.  Assuming a radiopurity for NaI crystals that is comparable to DAMA, the net predicted rate of events from the considered sources is 1.3-1.7\,cpd/kg/\keVee\ in the 1-10\,\keVee\ window where a WIMP signal is expected.   Although not shown in the tables, we also considered $^{60}$Co contamination of the steel vessel at the level of 7.2\,mBq/kg~\cite{SNOLAB}, which adds an additional 0.1-0.2\,cpd/kg/\keVee\ to the total.  The estimated background due to radioactivity is thus comparable to the total event rate observed by the DAMA collaboration.  Despite the assumption of high radiopurity, the NaI crystals themselves remain a large source of radioactivity.  Care must be taken to minimize this contribution as much as possible.  In this hypothetical setup, the steel vessel (Sandvik SANMAC SAF 2205 hollow bar duplex steel) is a significant component and can be further reduced by constructing the encapsulation from cleaner material or implementing a lining within the pressure vessel to shield the crystals.  
\begin{table}[thb]
\caption{Shown are the estimated contribution to event rate from 1-10\,\keVee\ in a single 8\ kg NaI crystal.  The first three items are calculated using the Geant4 simulation of a simple 8\ kg assembly.  The internal NaI contamination is taken from~\cite{vak2010} where the energy spectrum of events from radioactivity was simulated for the DAMA experiment assuming radioisotope concentrations reported in~\cite{dama-back}.}
\begin{center}
\begin{tabular}{lc} 
\hline
Material & event rate in NaI\\
         & (cpd/kg/\keVee)\\
\hline
drill ice                     & 0.8 \\
Antarctic ice                 & $<0.001$ \\
photomultiplier tubes         & 0.01-0.02 \\
steel PV                      & 0.2-0.6  \\
NaI crystal                   & $\sim$0.3\\
\hline
\end{tabular}
\end{center}
\label{tab:mc_output}
\end{table}

Neutrons, which will produce nuclear recoils in the NaI crystals, are also generated as a result of radioactivity through fission and $(\alpha,n)$ processes.  In these estimates, we neglect this source of background because the rate of neutrons generated by radioactivity is several orders of magnitude lower than the rate of gamma-rays.  Such events will thus be a small component of the net rate observed in the NaI scintillator.



\subsection{Background from Cosmic Rays}
\label{sec:cosmo_bg}

Cosmic muons produce background for dark matter experiments in the form of spallation products.  These consist of neutrons and excited nuclei, which may be long or short-lived.  Recently, long-lived phosphorescent states have also been postulated to be a product of cosmic activity in NaI crystals~\cite{Nygren}.  Owing to the large overburden achieved at underground laboratories, background produced by cosmic muons is expected to be well below the rate of events caused by radioactivity.  However, the cosmic muon flux, and thus the activity induced by it, is modulated with a period of about one year due to the change of temperature and pressure in the upper atmosphere.  Seasonal variations will thus play an important role in modulating this portion of the event rate and must be thoroughly understood.

It has been widely noted that the maximum of the DAMA signal (May 26$\pm$7 days~\cite{damafullpaper}) and the peak of the muon flux at LNGS (July 5$\pm$15 days~\cite{LVD}) bear a close proximity in time.   The amplitude of the muon modulation at the South Pole is 10\% and is strongly correlated with the atmospheric temperature.  The maximum muon rate occurs in mid-January~\cite{Tilav2010}. This modulation is larger than the 2\% effect at LNGS~\cite{LVD}.  Thus, any unremovable effects of cosmic ray origin that can be attributed to the DAMA signal will be opposite in phase and more pronounced at the South Pole.  Correlations between minor fluctuations in temperature and muon rate have also been observed and are well-characterized with IceCube data~\cite{Tilav2010}.  Such a high degree of correlation offers the opportunity to reject a cosmic-ray origin for any potential modulating signal.

It is worth noting that, at a depth of 2450\,m in the ice, the rate of muons and any associated spallation products will be comparable to the rate at deep underground sites.  Using MUSUN~\cite{musun}, the muon intensity at 2200 m.w.e. is estimated to be $1.3\times10^{-3}$\,${\rm \mu/m^2/s}$ with a mean energy of 307\,\gev.  Ice is more effective than rock at moderating low-energy ($<$10\,MeV) neutrons down to energies where they are unable to produce a nuclear recoil above threshold in a NaI detector.  Additionally, any remaining short-lived activity that happens to propagate into a crystal will occur in coincidence with a muon and thus can be vetoed.  IceCube has already demonstrated its utility as a robust atmospheric muon veto.  Within the volume of DeepCore, a low-energy extension to IceCube, a reduction of $8\times10^{-3}$ in muon rate has been achieved~\cite{DeepCoreICRC}.  Advanced software algorithms, currently being designed, will enhance the effectiveness by several orders of magnitude.


\section{Sensitivity}
\label{sec:sensitivity}

\begin{figure*}[!th]
\begin{center}
\begin{tabular}{cc}
\includegraphics[angle=0, width=0.46\textwidth]{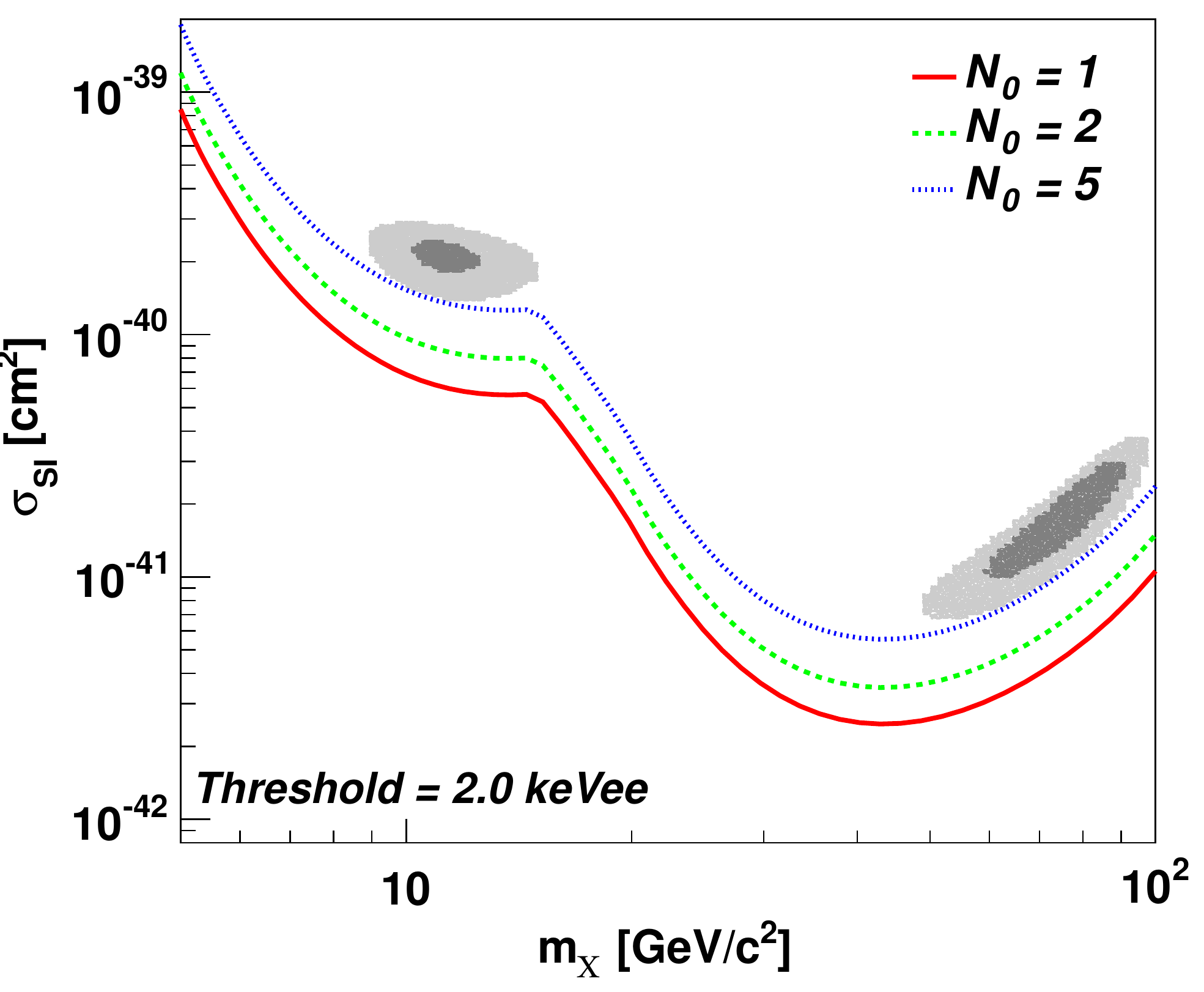} &
\includegraphics[angle=0, width=0.46\textwidth]{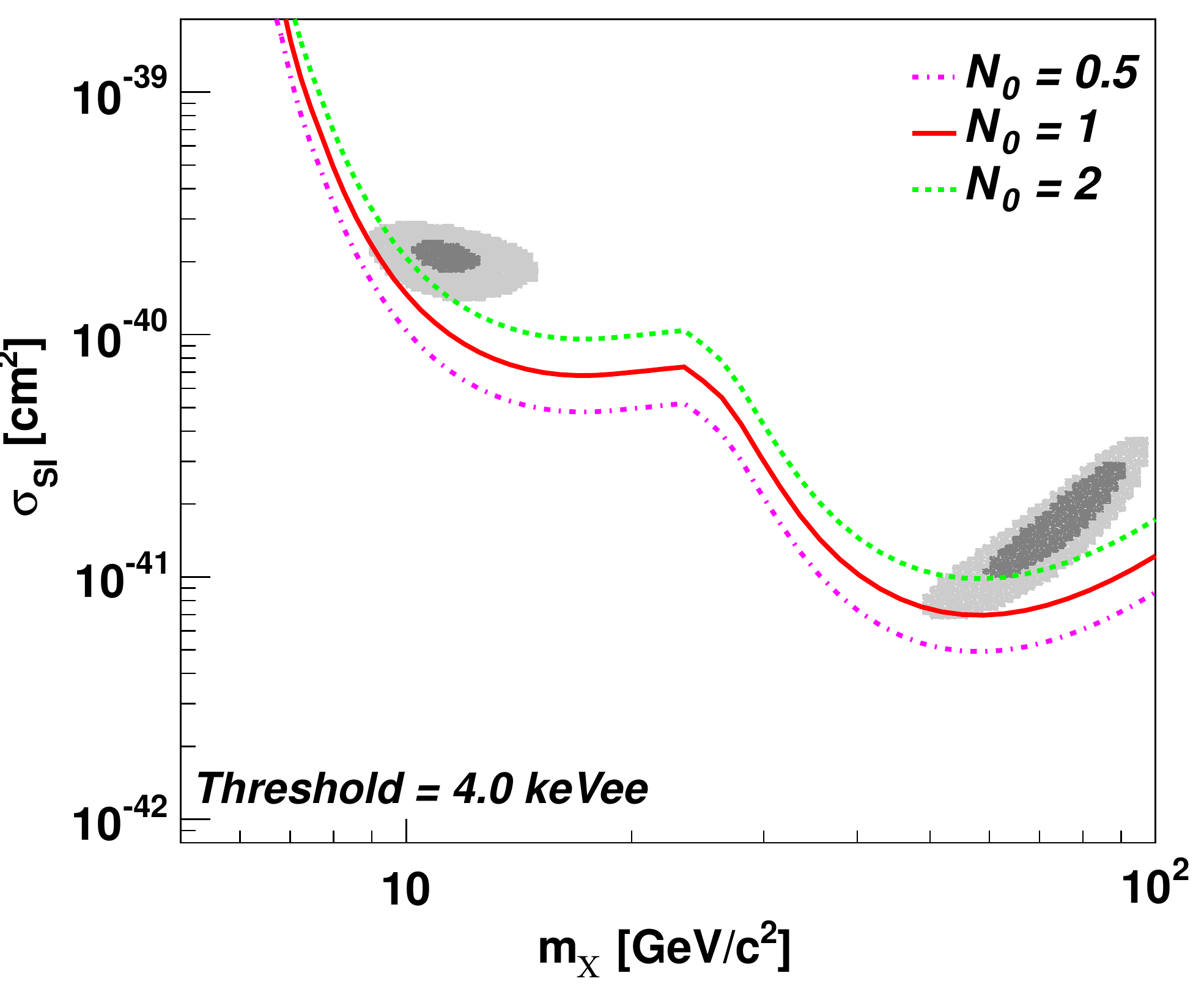} 
\end{tabular}
\caption{The curves show the sensitivity of hypothetical 500\,kg-year exposures with varying total event rates (in cpd/kg/\keVee).  Shown are two energy threshold scenarios.  The left plot shows sensitivities with a 2\,\keVee\ experimental threshold.  The right plot shows sensitivity with a 4\,\keVee\ threshold.  The gray regions show the 90\% (dark) and 99.7\% (light) DAMA/LIBRA allowed regions for interactions with Na (masses of $\sim$10\,\gev) and I (masses of $\sim$100\,\gev).  DAMA/LIBRA allowed regions are calculated without channeling.}
\label{fig:si_sensitivity_opt}
\end{center}
\end{figure*}


The DAMA/LIBRA and DAMA/NaI experiments are currently the world's most sensitive NaI-based dark matter experiments.  It is expected that an array of NaI detectors with the same mass, energy threshold and intrinsic background level as DAMA will achieve the same level of sensitivity as the DAMA experiments; however, it is less clear what level of sensitivity can be achieved with detectors that have a different reach in background level and energy threshold.   To explore these parameters, we calculate the sensitivity of a series of simulated experiments.

We consider an annual modulation arising from spin-independent elastic scattering of WIMPs, which are distributed according to the standard halo model, by following the prescription in~\cite{lewinSmith}.  Under these assumptions, the modulation of events at a specific recoil energy can be approximated with a cosine function as follows:
\begin{equation}
{\frac{dR(t,E)}{dE}} = {S_0(E)} + {S_m(E)} cos(\omega({t-t_c})),
\end{equation}
\label{eq:mod}
where $\omega = 2\pi/{T}$ and ${T} = 1~{\rm year}$.  The value $t_c$ is the time of year when the velocity of the WIMP halo relative to the Earth-bound experiment is at a maximum.  The energy, $E$, corresponds to the target recoil energy.  The sign of the modulation, \smod, can be positive or negative and depends on kinematic constraints.  \snaught\ is the unmodulated component of the dark matter signal.  

The DAMA collaboration analyze their data by measuring a residual rate of modulating events over the observed constant rate in a fixed interval of energy.  We follow the same procedure and integrate over the energy interval which corresponds to the region of experimentally accessible recoil energies:
\begin{equation}
{S_{0(m)} = \frac{\int_{E_{thresh}}^{E_{max}}{S_{0(m)}(E)\,dE} }{E_{max} - E_{thresh} }} .
\end{equation}
\label{eq:intRate}
This yields a simplified modulation equation:
\begin{equation}
{R(t)} = {S_0} + {S_m} cos(\omega({t-t_c})).
\end{equation}
\label{eq:modfit}
For the case of multiple target species in one detector, as with NaI, the net recoil rate is the sum of the recoil rates off the individual nuclei, weighted by the respective mass fractions. Because the time dependence remains the same for different targets, the simplified form for the time dependent rate shown in equation~\ref{eq:modfit} remains unchanged.

Experimental data are typically displayed using the calibrated electron-recoil equivalent energy (\keVee) because NaI cannot provide event-by-event discrimination between nuclear and electron recoils.  Working in the same units, we present results in two intervals in electron-recoil equivalent energy: [2, 6]\,\keVee\ and [4, 6] \keVee.  To convert the energy ranges defined in the experimentally measured units into recoil energy for the appropriate target nucleus, we use $q=0.3$\,\keVee/keV for Na recoils and $q=0.09$\,\keVee/keV for I recoils, where ${ E_{keVee} = qE_{recoil}}$ and $q$ is the quenching factor taken from~\cite{damaQuench}.  The energy threshold of 2\,\keVee\ is chosen based on the analysis thresholds and trigger efficiencies achieved by previous experiments such as DAMA/LIBRA~\cite{damafullpaper} and NAIAD~\cite{NAIAD}.  The higher 4\,\keVee\ threshold provides a comparison to a slightly more pessimistic scenario.  The upper bound is 6 \keVee.  Above this, the rate for recoils, in a standard halo model, drops well below the considered background rates of $0.5-5.0$\,cpd/kg/keVee.  In order to compare to DAMA data, we define the constant rate of events, \nzero, which is the sum of the time-independent dark matter component (\rzero) and the time-independent background.  

The sensitivity is calculated by fitting equation~\ref{eq:modfit} to a simulated dataset containing no modulation signal in it (null hypothesis).   The simulated data are created by randomly drawing events from a Poisson distribution with mean given by \nzero.   We report on several trials with \nzero\ ranging from 0.5 to 5 cpd/kg/\keVee.  The value $\nzero=0.5$ corresponds to an optimistic goal of achieving a better background rate compared to DAMA/LIBRA ($\nzero\approx1$).  The highest rate ($\nzero=5$) is characteristic of the background levels achieved by NAIAD~\cite{NAIAD}.  To obtain a representative statistical uncertainty, the total number of thrown events corresponds to the average number expected for a 250\,kg exposure over two years, integrated over the experimental energy range.  For a given value of \nzero, the simulated experiment is performed for the two energy intervals listed above.  

A simple chi-square fit allows ${S_0}$ and ${S_m}$ to float while keeping the frequency, $\omega$, and the phase, ${t_c}$, fixed.  Under the assumption of a null result, the 90\% C.L. upper limit to ${ S_m}$ is calculated from the error returned on the fit. The resulting sensitivities, in the form of 90\% C.L. upper limits to the spin independent cross-section,  are shown in Figure~\ref{fig:si_sensitivity_opt}.  For comparison, the calculated DAMA/LIBRA 90\% and 99.7\% allowed regions, without channeling~\cite{channeling}, are also displayed on the figures.

The results demonstrate that low energy thresholds of $\sim$2\,\keVee, or lower, are critical for obtaining adequate sensitivity to constrain the DAMA/LIBRA allowed regions.  Figure~\ref{fig:si_sensitivity_opt} shows that with a 2\,\keVee\ threshold, an experiment that has 5 times the DAMA event rate (and hence $\sim$5 times greater background) provides significant constraints with two years of exposure.  Considering the sensitivity curves for the higher threshold scenario, it is clear that control of background becomes more important.  With a 4\,\keVee\ energy threshold, one must improve upon the level of background achieved by DAMA/LIBRA in order to completely exclude the 3$\sigma$ allowed regions at 90\% C.L..  This result is understood because the expected rate of recoils fall off roughly exponentially with increasing recoil energy.  The feasibility of lowering the energy threshold of the experiment will depend on achieving a balance between increased signal and background rates.

Examining one set of sensitivity curves with varying background and fixed energy threshold, it is observed that the sensitivity for a given WIMP mass scales roughly inversely with the square root of \nzero.  This is easily understood because the sensitivity of the annual modulation search scales with the uncertainty in the number of unmodulated events, which is mostly background in this case.  This demonstrates that the modulation analysis implicitly performs a background subtraction.  For similar reasons, the sensitivity of a NaI experiment will increase with the square root of the exposure time or mass.

In this study, we have only considered sensitivities to spin-independent elastic scattering with a standard halo model. The ability to confirm or exclude a DAMA-like dark matter signal for alternative scenarios (e.g. spin-dependent scattering, inelastic models, etc.) is dependent on the kinematic constraints implied by the models.  Understanding the sensitivity reach for NaI under a variety of energy thresholds requires individual investigation.  The conclusions on relative sensitivity with scaling of background and exposure will remain largely the same.  As a model-independent exercise, we evaluate the statistical potential of a 2-year, 250\,kg exposure to confirm a DAMA-like signal without considering the nature of the possible underlying dark matter signal.  Events are drawn from a Poisson distribution assuming the same rate, phase and period of modulating events as observed in the DAMA/LIBRA 2-6 keVee energy region~\cite{Bernabei:2010mq}.   We find that the statistical power of the 250\,kg, two year, simulated dataset has a $>$5$\sigma$ significance and hence could confirm a DAMA-like signal to that level of significance.

\section{Summary \& Outlook}

In this paper, we have described a proposal for an annual modulation dark matter search in Antarctic ice using NaI scintillation crystals.   The ice provides a clean, quiet and stable environment, which is ideal for an annual modulation search.   Furthermore, seasonal effects on background will be out of phase with experiments in the Northern Hemisphere, thus providing an independent check of the DAMA results.   The sensitivity calculation presented in this work demonstrates that verification or exclusion of the DAMA signal can be achieved with a two year deployment of a 250\,kg experiment, assuming a comparable energy threshold and similar control of background as the DAMA experiments.

Preliminary simulation studies suggest that background from the NaI crystals and the materials used to instrument and encapsulate the crystals can be comparable or lower than background rates achieved by DAMA.  Radioactive background from crystals will remain a dominant component.  Future R\&D will focus on growing low-background crystals, designing low-background encapsulation for pressure shielding in the Antarctic ice, and developing a deployment plan.

A feasibility study on the remote operation of NaI crystals in South Pole ice is now underway.  Two crystals, totaling 17\,kg, were tested at the Boulby Underground Laboratory and later moved to the University of Wisconsin at Madison where they were encased in stainless steel vessels.  During the polar season 2010/2011, the units were attached to IceCube strings and successfully deployed to a depth of 2450\,m in the ice.  Data are being gathered from the prototypes.  Results from this demonstration will be available in an ensuing publication.  

\section{Acknowledgements}

The concept of a dark matter annual modulation experiment in the Southern Hemisphere has been discussed by several people in the past.  For the development of the experiment presented here, we wish to thank Peter Fisher, Bernard Sadoulet and Christopher Stubbs for their encouragement and input. We acknowledge Katherine Freese for useful theoretical discussion.  We wish to thank Jonghee Yoo for sharing the Pipewimp limit calculation package, which was used to perform the sensitivity studies presented in this paper.  We thank SNOLAB for their efforts on the low background measurements.  We thank STFC and the Boulby mine company CPL for support of low background measurements of NaI.  This work was supported by the IceCube Research Center, the Wisconsin Alumni Research Foundation, the Sloan Research Foundation, the National Science and Engineering Research Council of Canada, the University of Alberta, Science Undergraduate Laboratory Internships and Fermilab, which is operated by Fermi Research Alliance, LLC under Contract No. De-AC02-07CH11359 with the United States Department of Energy.







\end{document}